\documentclass[10pt]{article}

\usepackage[utf8]{inputenc}
\usepackage[T1]{fontenc}
\usepackage{amssymb,amsmath,dsfont}
\usepackage{graphicx}
\usepackage[usenames,dvipsnames]{color}
\usepackage[notref,notcite,final]{showkeys}
\usepackage[pagewise,displaymath,modulo]{lineno}
\usepackage[left=2cm,top=1cm,right=3cm,bottom=3cm]{geometry} 
\usepackage{url} 

\usepackage[round]{natbib}
\author{Stefan Siegert\footnote{Corresponding author's address: University of Exeter, Harrison Building, North Park Road, Exeter, EX4 4QF, United Kingdom. Email: s.siegert@exeter.ac.uk}}
\title{Variance estimation for Brier Score decomposition\footnote{Submitted to {\it Quarterly Journal of the Royal Meteorological Society}}}
\date{Draft version: \today}

\begin{document}
\twocolumn
\maketitle

\begin{abstract}

The Brier Score is a widely-used criterion to assess the quality of probabilistic predictions of binary events. 
The expectation value of the Brier Score can be decomposed into the sum of three components called reliability, resolution, and uncertainty which characterize different forecast attributes.
Given a dataset of forecast probabilities and corresponding binary verifications, these three components can be estimated empirically.
Here, propagation of uncertainty is used to derive expressions that approximate the sampling variances of the estimated components.
Variance estimates are provided for both the traditional estimators, as well as for refined estimators that include a bias correction.
Applications of the derived variance estimates to artificial data illustrate their validity, and application to a meteorological prediction problem illustrates a possible use case.
The observed increase of variance of the bias-corrected estimators is discussed.

\end{abstract}

\newcommand{\br}{\text{Br}}
\newcommand{\rel}{\text{REL}}
\newcommand{\res}{\text{RES}}
\newcommand{\unc}{\text{UNC}}
\newcommand{\indic}{\mathds{I}}
\newcommand{\one}{\mathbf{1}}
\newcommand{\e}{\mathbb{E}}
\newcommand{\vv}{\mathbb{V}}
\newcommand{\jac}{\mathcal{D}}
\newcommand{\unity}{\mathbf{I}}
\newcommand{\covv}{\mathbb{C}\text{ov}}
\newcommand{\bu}{\text{\tiny\textbullet}}
\newcommand{\bx}{\mathbf{x}}
\newcommand{\bd}{\mathbf{d}}
\newcommand{\bc}{\mathbf{c}}
\newcommand{\refsec}[1]{Sec.~\ref{#1}}
\newcommand{\refSec}[1]{Section~\ref{#1}}
\newcommand{\refeq}[1]{Eq.~(\ref{#1})}
\newcommand{\refEq}[1]{Equation~(\ref{#1})}
\newcommand{\reftab}[1]{Table~\ref{#1}}
\newcommand{\refTab}[1]{Table~\ref{#1}}
\newcommand{\reffig}[1]{Fig.~\ref{#1}}
\newcommand{\refFig}[1]{Figure~\ref{#1}}
\newcommand{\refapp}[1]{Appendix~\ref{#1}}

\section{Introduction}

The basis of the following discussion is a data set of forecast probabilities $\{p_n\}_{n=1}^N$, and corresponding verifications $\{y_n\}_{n=1}^N$.
A binary prediction setting is assumed, that is, the verification at instance $n$, $y_n$, is either one if the event happens, or zero if it does not happen.
The forecast probability $p_n$ is a probabilistic prediction for the event $y_n=1$.
The empirical Brier Score \citep{brier1950verification} assigned to the set of forecasts $\{p_n\}$ is given by
\begin{equation}
\br = \frac{1}{N}\sum_{n=1}^N (p_n - y_n)^2.\label{eq185}
\end{equation}

The Brier Score is negatively oriented, assigning lower values to better forecasts.
The Brier Score further has the property of being {\it proper}, which means that a forecaster cannot improve his expected Brier Score by issuing forecasts $q$ that differ from his best estimates $p$ of the actual event probabilities.
In fact, any such deviance from $p$ will increase his expected Brier Score, which makes the Brier Score a {\it strictly proper} scoring rule \citep{degroot1983comparison}.

It has been shown by \citet{murphy1973new} that the Brier Score can be decomposed additively into three non-negative terms, called reliability, resolution, and uncertainty:
\begin{equation}
    \br=\rel-\res+\unc.\label{eq932}
\end{equation}
A qualitative interpretation of the individual components is given next; mathematical details follow below.
The reliability term quantifies how far the forecast probabilities $p_n$ differ from the corresponding conditional event probabilities $\mathds{P}(y_n=1\mid p_n)$. 
Ideally, it should always hold that $p_n = \mathds{P}(y_n=1\mid p_n)$; in this case the reliability component vanishes.
A systematic difference between the two terms is penalized by a positive reliability component.
The resolution component rewards variations of the forecast probabilities that are consistent with varying event probabilities.
A forecasting scheme that constantly issues the same probabilities has zero resolution.
Any meaningful variability of the forecast leads to a positive resolution term which improves the Brier Score.
The uncertainty component is equal to the Brier Score of the average (climatological) probability.
It thus serves as a benchmark to which the Brier Score of the forecast under consideration can be compared.
A `useful' forecast should have a Brier Score that is lower than its uncertainty component, or in other words, the resolution should be larger than the reliability.

Consider the forecast probability $p$ and the corresponding verification $y$ as two (dependent) random quantities.
Then the {\it calibration function} $\pi(p)$ and the {\it climatology} $\bar{\pi}$ are defined as
\begin{align}
\pi(p) & = \mathds{P}(y=1\mid p)\text{, and}\\
\bar{\pi} &= \mathds{P}(y=1).
\end{align}
Using these definitions, the three components of the Brier Score decomposition are formally given by
\begin{align}
\rel^* &= \e\left[p-\pi(p)\right]^2,\label{eq123}\\
\res^* &= \e\left[\pi(p)-\bar{\pi}\right]^2\text{, and}\label{eq124}\\
\unc^* &= \bar{\pi}(1-\bar{\pi}),\label{eq125}
\end{align}
where $\e$ denotes the mathematical expectation value \citep{broecker2009reliability}.
The star ($^*$) is used to differentiate the exact analytical expressions from their empirical estimators, which are discussed below.

In practice, the three components of the Brier Score decomposition must be estimated empirically from the set of forecast probabilities and corresponding verifications $\{p_n,y_n\}$.
Such estimators are derived in \citet{murphy1973new}; they are presented below in a somewhat different notation, which is suitable for variance estimation by propagation of uncertainty (see \refsec{sec20}).

First of all, the observed forecast probabilities $\{p_n\}$ are binned into $D$ mutually exclusive and collectively exhaustive bins $p_\sqcup^d$, where $d=1,\cdots,D$.
Here, bins of equal width which are half-open to the left are used, except the first bin which is closed (but the theory also applies to variable bin widths).
As an example, if $D=3$ one would have $\{p_\sqcup^d\}_{d=1}^3=\{[0,1/3],(1/3,2/3],(2/3,1]\}$.
Using this binning of the forecast probabilities, the following matrices are defined:
\begin{align}
A \in \{0,1\}^{N\times D}:\ & A_{nd} = \indic(p_n \in p_\sqcup^d), \label{eqn05}\\
B \in \{0,1\}^{N\times D}:\ & B_{nd} = \indic(p_n \in p_\sqcup^d)\ y_n, \label{eqn06}\\
C \in [0,1]^{N\times D}:\ & C_{nd} = \indic(p_n \in p_\sqcup^d)\ p_n,\label{eqn07}\\
Y \in \{0,1\}^{N\times 1}:\ &  Y_n = y_n,\label{eqn08}
\end{align}
where $\indic(\cdot)$ denotes the indicator function.
Summation over a column or row of a matrix is abbreviated by a bullet ($\bu$), for example
\begin{equation}
A_{\bu d} = \sum_{n=1}^N A_{nd}.
\end{equation}
A bullet without a second index always refers to the row vector of column sums of a matrix, as in
\begin{equation}
A_\bu = \one^T A
\end{equation}
where $\one$ is the $N\times 1$ column vector with all elements equal to one.

Using these definitions, $A_{\bu d}$ is equal to the total number of cases where $p_n \in p_\sqcup^d$.
$B_{\bu d}$ is equal to the number of cases where $p_n \in p_\sqcup^d$ and at the same time $y_n=1$.
Therefore, a binned estimator for the calibration function is given by
\begin{equation}
    \pi(p) \approx \pi_d = \frac{B_{\bu d}}{A_{\bu d}},
\end{equation}
where $p \in p_\sqcup^d$. The climatology is estimated by
\begin{equation}
\bar{\pi} \approx \frac{Y_\bu}{N}.
\end{equation}
Furthermore, $C_{\bu d}/A_{\bu d}$ is equal to the average forecast probability in the $d$-th bin.
$Y_\bu$ is equal to the total number of events that have occurred.
Lastly, note that $B_{\bu\bu} = Y_\bu$, and $A_{\bu\bu} = N$.

The above notation is closely related to the {\it contingency table} $N_{kd}$, defined by
\begin{equation}
    N_{kd} = \sum_{n=1}^N \indic(y_n=k)\indic(p_n\in p_\sqcup^d),
\end{equation}
which is commonly used to estimate Brier Score decompositions of binned probabilistic forecasts.
The matrices $A$, $B$, $C$, and $Y$ were introduced to facilitate variance estimation.

Using the matrices $A$, $B$, $C$, and $Y$, the estimators for the three components of the Brier Score decomposition originally proposed by \cite{murphy1973new} are given by
\begin{align}
\rel&=\rel(A_\bu,B_\bu,C_\bu)\nonumber\\
&=\frac{1}{N}\sum_{d\in\mathds{D}_0} \frac{1}{A_{\bu d}}\left(B_{\bu d} - C_{\bu d}\right)^2,\label{eq77}\\ 
\res&=\res(A_\bu,B_\bu,Y_\bu)\nonumber\\
&=\frac{1}{N}\sum_{d\in\mathds{D}_0} A_{\bu d}\left(\frac{B_{\bu d}}{A_{\bu d}} - \frac{Y_\bu}{N}\right)^2\\ 
\unc&=\unc(Y_\bu)\nonumber\\
&=\frac{Y_\bu(N-Y_\bu)}{N^2},
\end{align}
where $\mathds{D}_0=\{d:A_{\bu d}>0\}$.
In the following $\rel$, $\res$, and $\unc$ are referred to as the {\it traditional estimators} of the components of Brier Score decomposition.

In \cite{ferro2012unbiased} it is shown that the traditional estimators are biased. They show that the bias can be corrected to some extent, although never perfectly eliminated. Using the above notation, the estimators proposed by \cite{ferro2012unbiased} are given by
\begin{align}
&\rel'(A_\bu, B_\bu, C_\bu)\nonumber\\
& = \rel - \frac{1}{N}\sum_{d\in\mathds{D}_1}\left\{\frac{B_{\bu d}(A_{\bu d}-B_{\bu d})}{A_{\bu d}(A_{\bu d}-1)}\right\},\label{eq317}
\end{align}
\begin{align}
\res'(A_\bu,B_\bu,&Y_\bu) \nonumber\\
=\res -&\frac{1}{N}\sum_{d\in\mathds{D}_1}\left\{\frac{B_{\bu d}(A_{\bu d}-B_{\bu d})}{A_{\bu d}(A_{\bu d}-1)}\right\}\nonumber\\
 +&\frac{Y_\bu(N-Y_\bu)}{N^2(N-1)},
\end{align}
and
\begin{align}
\unc'(Y_\bu)&=\unc + \frac{Y_\bu(N-Y_\bu)}{N^2(N-1)}\nonumber\\
&=\frac{Y_\bu(N-Y_\bu)}{N(N-1)},\label{eq319}
\end{align}
where $\mathds{D}_1=\{d:A_{\bu d}>1\}$. 
$\rel'$, $\res'$, and $\unc'$ are referred to as the {\it bias-corrected estimators}.

Due to the analytical expressions \refeq{eq123} -- \refeq{eq125}, it holds that $\rel^*\in[0,1]$, $\res^*\in[0,1]$ and $\unc^*\in[0,0.25]$.
One could argue that estimators for the individual components should be confined to these intervals as well.
While the traditional estimators always satisfy this restriction, the bias-corrected estimators do not.
\citet{ferro2012unbiased} acknowledge the possibilities $\rel'<0$ and $\res'<0$, and recommend a suitable modification to their bias correction.
Unfortunately, this modification does not account for the possibilities $\unc'>0.25$ and $\res'>1$.
In \refapp{secAppAvoid} a modification of the bias-corrected estimators is suggested which avoids all possible inconsistencies.

Note that binning of continuous forecast probabilities for Brier Score decomposition introduces an estimation error which can lead to the effect that the estimated components do not add up to the average Brier Score calculated by \refeq{eq185}.
Two additional components related to within-bin variance and within-bin covariance have been introduced to account for this effect \citep{stephenson2008extra}. 
Furthermore, binning of forecast probabilities is not the only possible method to estimate the calibration function.
\citet{broecker2008remarks} uses kernel density estimation and chooses the kernel bandwidth by leave-one-out Brier Score minimization.
The latter method can also be applied to choose the number of bins in the binning-and-counting approach.
In order to maintain a certain level of simplicity, the above refinements of Brier Score decomposition are not considered in the present paper.

A note on terminology: 
In order to limit confusion due to repeated use of the word {\it estimate}, we shall always use the term {\it estimator} to refer to the components of the Brier Score decomposition estimated by \refeq{eq77} -- \refeq{eq319}, and the term {\it variance estimates} to refer to the approximated variance of these components.

In \refsec{sec20} of this article it is shown how propagation of uncertainty can be applied to calculate variance estimates for the estimators of a Brier Score decomposition.
The variance estimates are validated in an artificial prediction setting in \refsec{secExample1}.
Application to a meteorological prediction problem in \refsec{secExample2} illustrates a possible use case.
In \refsec{secDisc} the simplifying assumptions, validity of the new variance estimates, and variance increase of the bias-corrected estimators are discussed.
\refSec{secSummary} concludes the article. 
The article is complemented with Supplementary Online Material which includes source code written in the {\tt R} programming environment \citep{Rmanual2012} to reproduce all calculations. 
A library for the {\tt R} environment \citep{bride2013} is available to apply the results of this study in practice.

\section{Variance estimation by propagation of uncertainty}\label{sec20}

The general setting is now that we have scalar estimators $F$ for the components of a Brier Score decomposition, which depend nonlinearly on the column sums $\bx$ of a matrix $X$.
%
%
For example if $F=\rel$ we have 
\begin{align}
X &=[A|B|C]\in \mathds{R}^{N\times 3D}\\
\bx & = \one^T X = [A_\bu | B_\bu | C_\bu] \in \mathds{R}^{1\times 3D}.
\end{align}
It is possible to apply {\it propagation of uncertainty} \citep[e.~g.][]{mood1974introduction} to estimate the variance of $F(\bx)$ as a function of the covariances of its arguments.
The first-order Taylor expansion of $F$ around $\bar{\bx}$ (the expectation value of $\bx$) is given by
\begin{equation}
F(\bx) \approx F(\bar{\bx}) + \frac{\partial F(\bar{\bx})}{\partial\bx} (\bx-\bar{\bx})^T,\label{eqn79}
\end{equation}
where $\partial F(\bar{\bx})/\partial \bx$ is shorthand for the Jacobian of $F(\bx)$ evaluated at $\bar{\bx}$.
Under this approximation, the variance of $F(\bx)$ is given by
\begin{align}
\vv[F(\bx)]& = \e[F(\bx)-\e F(\bx)]^2\\
& = \frac{\partial F(\bar{\bx})}{\partial \bx} \covv(\bx) \frac{\partial F(\bar{\bx})}{\partial\bx^T},\label{eq805}
\end{align}
where $\covv(\bx) =\e[(\bx-\bar{\bx})^T (\bx -\bar{\bx})]$.
Recall that the $i$-th element of $\bx$ is the sum over the $i$-th column of $X$.
Under the assumption that the rows of $X$ are independent and identically distributed, it can be shown that 
\begin{align}
\covv(\bx) \approx X^T \left(\unity - \frac{1}{N}\one \one^T \right) X,\label{eq492}
\end{align}
where $\unity$ denotes the identity matrix. 
\refEq{eq492} is derived using the fact that $\covv(\bx_i,\bx_j)$ is equal to $N$ times the covariance between the elements of the $i$-th and $j$-th column of the matrix $X$.

\refEq{eq805} combined with \refeq{eq492} provides a recipe to estimate the variances of the estimators $\rel$, $\res$, and $\unc$, as well as their bias-corrected counterparts.
All data that is necessary to estimate the variances has already been calculated for the estimators themselves.
The only tedious bit is the calculation of the derivatives of the estimators with respect to the individual column sums for the Jacobian.
These derivatives are given in \refapp{secAppDeriv}.

\section{Application to artificial data}\label{secExample1}

In order to illustrate their validity, the variance estimates are applied to Brier Score decomposition in an artificial prediction setting, for which the components of the decomposition are known analytically.
The results are discussed in \refsec{secDisc}.
The code to reproduce the numerical computations of this article is available in the Supplementary Online Material.

In the artificial example, assume that the event $y\in\{0,1\}$ is an independent realization of a Bernoulli trial with success probability $q$.
The value $y=1$ means that `the event occurs'.
In the present example, the event probability $q$ is itself a random variable that is equally likely to assume one of 6 possible values, namely $q\in\{q_d\}_{d=1}^6=\{0.05,0.15,\cdots,0.55\}$.
A forecasting scheme for the event $y$ which has nonvanishing resolution and nonvanishing reliability is constructed as follows:
The forecast probability $p$ corresponds to the actual event probability $q$ whenever $q\neq 0.55$.
But whenever the event probability $q=0.55$, the forecast probability is equal to $p=1$.
That is, $p\in\{p_d\}_{d=1}^6=\{q_1,\cdots,q_5,1\}$, with equal probability of $\frac{1}{6}$.

For the above scheme, the climatological probability is equal to 
\begin{align}
\bar{\pi}&=\frac{1}{6}\sum_{d=1}^6 q_d=\frac{3}{10}.
\end{align} 
The true uncertainty of this forecasting scheme is thus given by 
\begin{align}
\unc^*=\bar{\pi}(1-\bar{\pi})=\frac{21}{100}.
\end{align}
Furthermore, since the calibration function in this setting is given by
\begin{equation}
\pi(p_d)=q_d,
\end{equation}
the true reliability component of the Brier Score of the forecast $p$ is calculated as
\begin{align}
\rel^* = \frac{1}{6}\sum_{d=1}^6 (p_d - q_d)^2 = \frac{27}{800},
\end{align}
and the true resolution of the forecast is given by
\begin{align}
\res^* = \frac{1}{6}\sum_{d=1}^6 (q_d-\bar{\pi})^2 = \frac{7}{240}.
\end{align}
Note that in this example $\rel^* > \res^*$, and therefore the forecast is `useless' in the sense that the constant climatological probability $\bar{\pi}$ achieves a better Brier Score (which is equal to $\unc^*$) than the forecast probabilities $p$.

A single numerical experiment consists of $N=250$ forecast probabilities $p_n$, and corresponding event indicators $y_n$, independently sampled as outlined above.
Each such experiment results in a data set of forecasts and verifications $\{p_n,y_n\}_{n=1}^N$, and a Brier Score decomposition is estimated for this data set.
The forecast probabilities assume only 6 discrete values, and therefore no artificial binning has to be applied.
For infinitely many forecast instances the estimators converge to the true components, without discrepancies introduced by the binning.
The resulting estimators $\rel$, $\res$, and $\unc$, as well as their bias-corrected counterparts $\res'$, $\rel'$, and $\unc'$ are calculated for this data, together with their corresponding variance estimates derived in \refsec{sec20}.
This whole experiment is repeated 100 times, each time with a new realization of forecast probabilities $p_n$ and corresponding event indicators $y_n$.

\begin{figure*}
\centering
\includegraphics{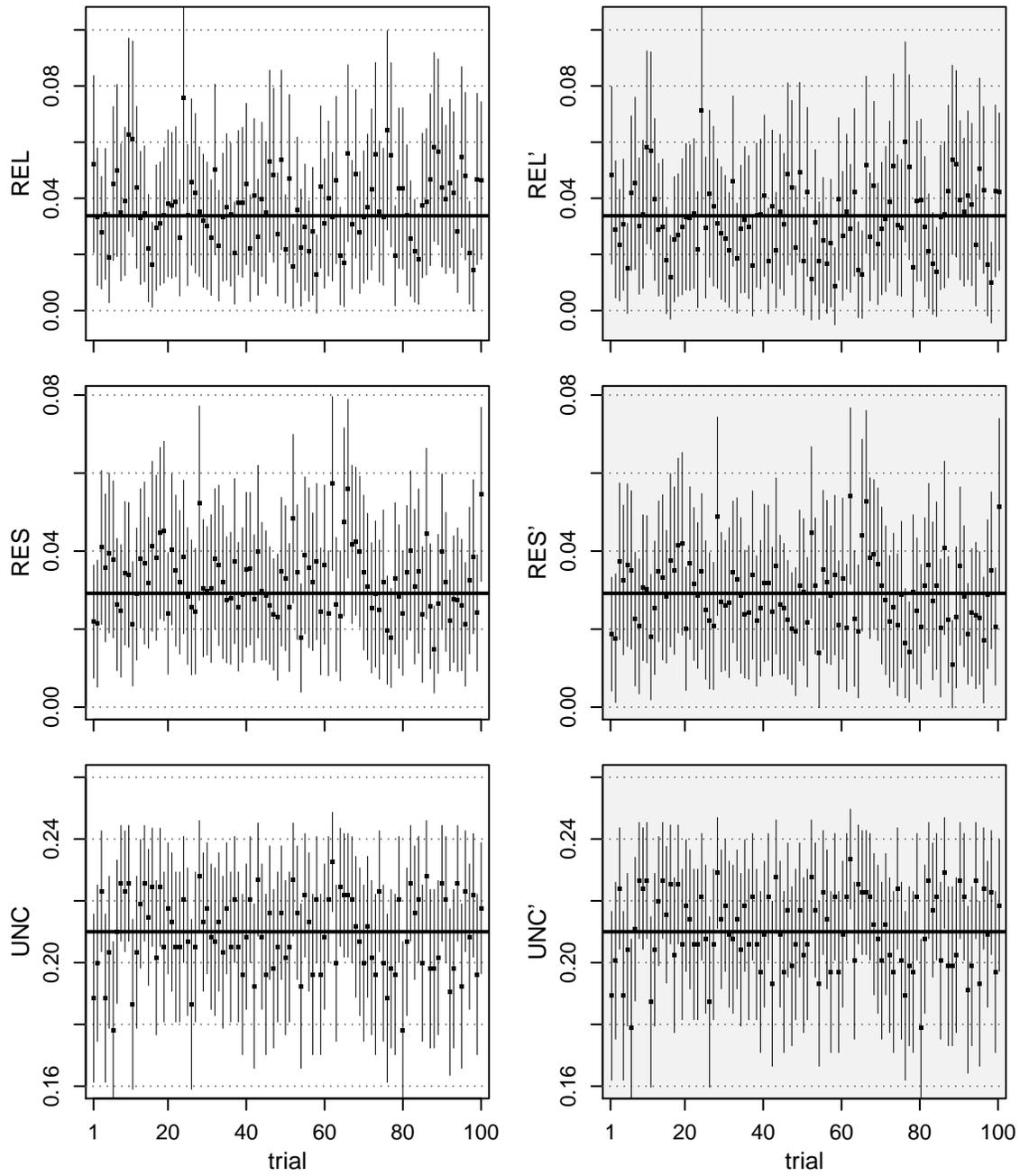}
\caption{Illustration of the experiment with artificial data.
For each trial of the experiment, the traditional and bias-corrected estimators of the Brier Score components are shown, augmented with error bars with a half-width of $2$ estimated standard deviations.
The bold black line indicates the true value.
All abscissae are as in the lower panels.}
\label{fig1}
\end{figure*}

The results of these 100 trials are illustrated in \reffig{fig1}.
For each trial, the traditional (left) and bias-corrected (right) estimators for reliability, resolution, and uncertainty are shown, augmented with error bars with a half width of two estimated standard deviations.

\begin{table*}
\caption{Summary of the artificial example. All averages are taken over the 100 trials of \reffig{fig1}. The first column shows the sample variance of the estimators. The second column shows the average of the estimated variances. The third column shows the average squared difference between the estimator and the true value. The fourth column shows the average bias, that is the average difference between the estimated value and the true value.}
\centering
\begin{tabular}{lrrrr}
& sample variance & avg. est. variance & avg. squared error & avg. bias \\
\hline
$\rel$ & $1.540 \times 10^{-4}$ & $1.665 \times 10^{-4}$ & $1.641 \times 10^{-4}$ & $3.182 \times 10^{-3}$ \\ 
$\rel'$ & $1.548 \times 10^{-4}$ & $1.663 \times 10^{-4}$ & $1.563 \times 10^{-4}$ & $-1.184 \times 10^{-3}$ \\ 
$\res$ & $7.062 \times 10^{-5}$ & $8.521 \times 10^{-5}$ & $8.093 \times 10^{-5}$ & $32.101 \times 10^{-4}$ \\ 
$\res'$ & $7.220 \times 10^{-5}$ & $8.776 \times 10^{-5}$ & $7.230 \times 10^{-5}$ & $-3.155 \times 10^{-4}$ \\ 
$\unc$ & $1.561 \times 10^{-4}$ & $1.336 \times 10^{-4}$ & $1.565 \times 10^{-4}$ & $-6.195 \times 10^{-4}$ \\ 
$\unc'$ & $1.573 \times 10^{-4}$ & $1.347 \times 10^{-4}$ & $1.574 \times 10^{-4}$ & $2.214 \times 10^{-4}$ \\ 
\end{tabular}

\label{tab1}
\end{table*}

In \reftab{tab1}, the outcome of the experiment is further quantified by statistical summary measures.
To make the calculation of these summary measures precise, consider as an example the estimator $\rel$. 
Define $\rel_i$ to be the estimator $\rel$ obtained on the $i$-th trial and $\vv\rel_i$ the corresponding estimated variance. 
Furthermore define $\overline{\rel}=\frac{1}{100}\sum_{i=1}^{100}\rel_i$. 
The sample variance (first column) was calculated by $\frac{1}{100}\sum_{i=1}^{100}(\rel_i-\overline{\rel})^2$, the average estimated variance (second column) was calculated by $\frac{1}{100}\sum_{i=1}^{100}\vv\rel_i$, the average squared error (third column) was calculated by $\frac{1}{100}\sum_{i=1}^{100}(\rel_i-\rel^*)^2$, and the average bias (fourth column) was calculated by $\frac{1}{100}\sum_{i=1}^{100}(\rel_i-\rel^*)$.
Summary measures for the other components were calculated accordingly.

The artificial example of the present section was used to check convergence rates of the variance estimates.
The average absolute difference between sample variance and estimated variance decays proportional to $N^{-3/2}$.
This result holds for all components of the Brier Score decomposition.
However, it was only obtained empirically, and might not generalize to different prediction settings.
A rigorous proof of estimator convergence rates is outside the scope of this paper.

\section{Meteorological application}\label{secExample2}

In this section, Brier Score decomposition is applied to real forecast data and the variance estimates are used to quantify the variability of the components of the decomposition.
We use daily maximum temperature observations measured at Dresden/Germany (WMO no. 10488) between 1 January 1980 and 31 December 1999 \citep{dwd2012}.
The (binary) prediction target is the exceedance of a certain threshold one day in the future.

The data between 1 January 1980 and 31 December 1989 is used as training data.
Denote this data by $T'_n$, where $n$ is an integer that indicates `days since 1 January 1970'.
We omit the unit of $T'_n$ and remember that it is measured in $^\circ$C.
The {\it seasonal cycle} $c_n$ is obtained by fitting a second order trigonometric polynomial to the observations:
\begin{align}
c_n = \beta_0 & + \beta_1 \cos(\omega n) + \beta_2 \sin(\omega n) \nonumber\\
&+ \beta_3 \cos(2 \omega n) + \beta_4 \sin(2 \omega n),
\end{align}
where $\omega=2\pi/(365.2425\text{ days})$ and the coefficients were fitted by minimizing the sum of squared differences between $c_n$ and $T'_n$ using ordinary linear regression. For the data at hand, we obtain 
\begin{equation}
\{\beta_0,\cdots,\beta_4\}=\{13.2, -10.7, -3.1, -0.6, 0.03\}\nonumber
\end{equation}
over the training period. Using the seasonal cycle, the {\it anomalies} $T_n$ are defined by
\begin{equation}
T_n = T'_n - c_n.
\end{equation}
The anomalies in the training period have zero mean (by construction) and a standard deviation of $4.7$.
Next, a first-order autoregressive (AR1) model is fitted to the anomalies, using the {\tt R} function {\tt ar} provided by the {\tt stats} package \citep{Rmanual2012}. That is, the temperature anomaly $T_{n+1}$, conditional on the anomaly $T_n$ is modeled by
\begin{equation}
T_{n+1} = \alpha T_n + \sigma \epsilon_n,
\end{equation}
where $\alpha$ is the AR1 parameter which quantifies the serial dependence of successive temperature anomalies, $\sigma^2$ is the variance of the residuals, and $\epsilon_n$ is a realization of unit variance Gaussian white noise.
The values $\alpha=0.77$ and $\sigma=2.97$ are obtained in the training data.

The prediction target is whether the temperature anomaly at time $n$ exceeds a threshold $\tau^\circ$C on the next day, that is $y_n = \indic(T_n > \tau)$.
Using the autoregressive model, a probabilistic 24h exceedance forecast is produced using the formula
\begin{equation}
p_n = \mathds{P}(T_n > \tau \mid T_{n-1} = t) = 1-\Phi_{\alpha t,\sigma}(\tau),\label{eq173}
\end{equation}
where $\Phi_{\mu,\sigma}(x)$ is the cumulative Gaussian distribution function with mean $\mu$ and variance $\sigma^2$, evaluated at $x$.
Using \refeq{eq173} and the parameters obtained from the training data, daily forecasts are produced for the time between 1 January 1990 and 31 December 1999.
The forecast probabilities $p_n$ for the targets $y_n$ are analyzed by decomposition of the Brier Score, with the number of equidistant bins set to $D=10$.

\begin{table*}
\caption{Summary of Brier Score decomposition of 10 years' worth of temperature anomaly exceedance forecasts (1 day lead time, threshold $5^\circ$C) by an autoregressive model.}\label{tab2}
\centering
\begin{tabular}{ccccccc}
& REL & RES & UNC & REL$'$ & RES$'$ & UNC$'$ \\ 
\hline
estimate  & $ 9.060 \times 10^{-4} $ & $ 0.0542 $ & $ 0.1408 $ & $ 4.130 \times 10^{-4} $ & $ 0.0537 $ & $ 0.1408 $\\ 
variance  & $ 2.096 \times 10^{-7} $ & $ 1.157 \times 10^{-5} $ & $ 1.684 \times 10^{-5} $ & $ 2.056 \times 10^{-7} $ & $ 1.164 \times 10^{-5} $ & $ 1.685 \times 10^{-5} $\\ 
\end{tabular}

\end{table*}

The result of the analysis is presented in \reftab{tab2} for the choice of the threshold $\tau=5$.
Estimators of the three components $\rel$, $\res$, and $\unc$, in the traditional and the bias-corrected version are given in the first row.
Both, the traditional and bias-corrected estimators add up to $\rel-\res+\unc=0.0875$.
The empirical Brier Score calculated by \refeq{eq185} is equal to $\br=0.0868$.
The difference between $\br$ and $\rel-\res+\unc$ is due to the binning of the continuous AR1 forecast probabilities, which was applied to estimate the decomposition.
In the second row of \reftab{tab2}, the corresponding variance estimates are shown.

\begin{figure}
\centering
\vspace{1cm}
\includegraphics[width=.45\textwidth]{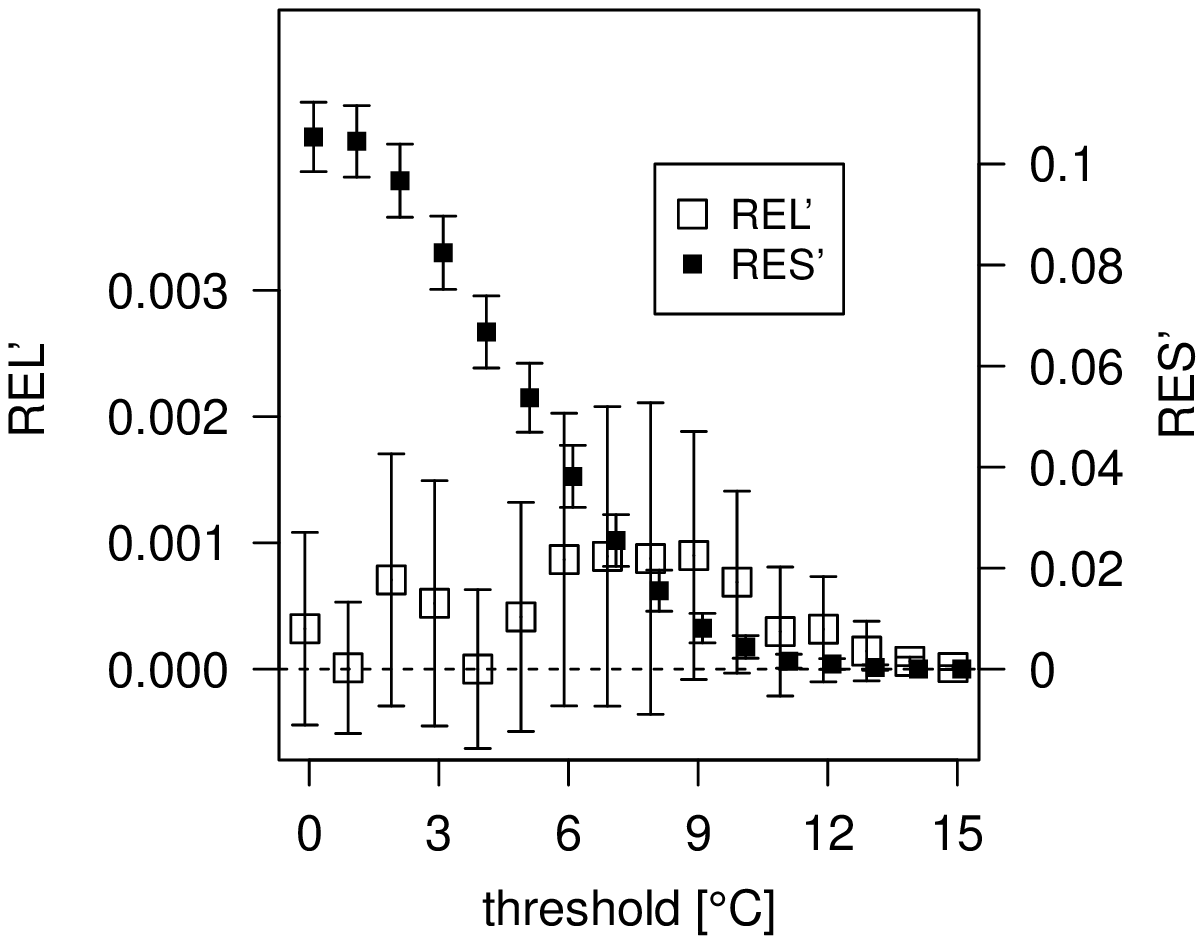}\\[5mm]
\includegraphics[width=.45\textwidth]{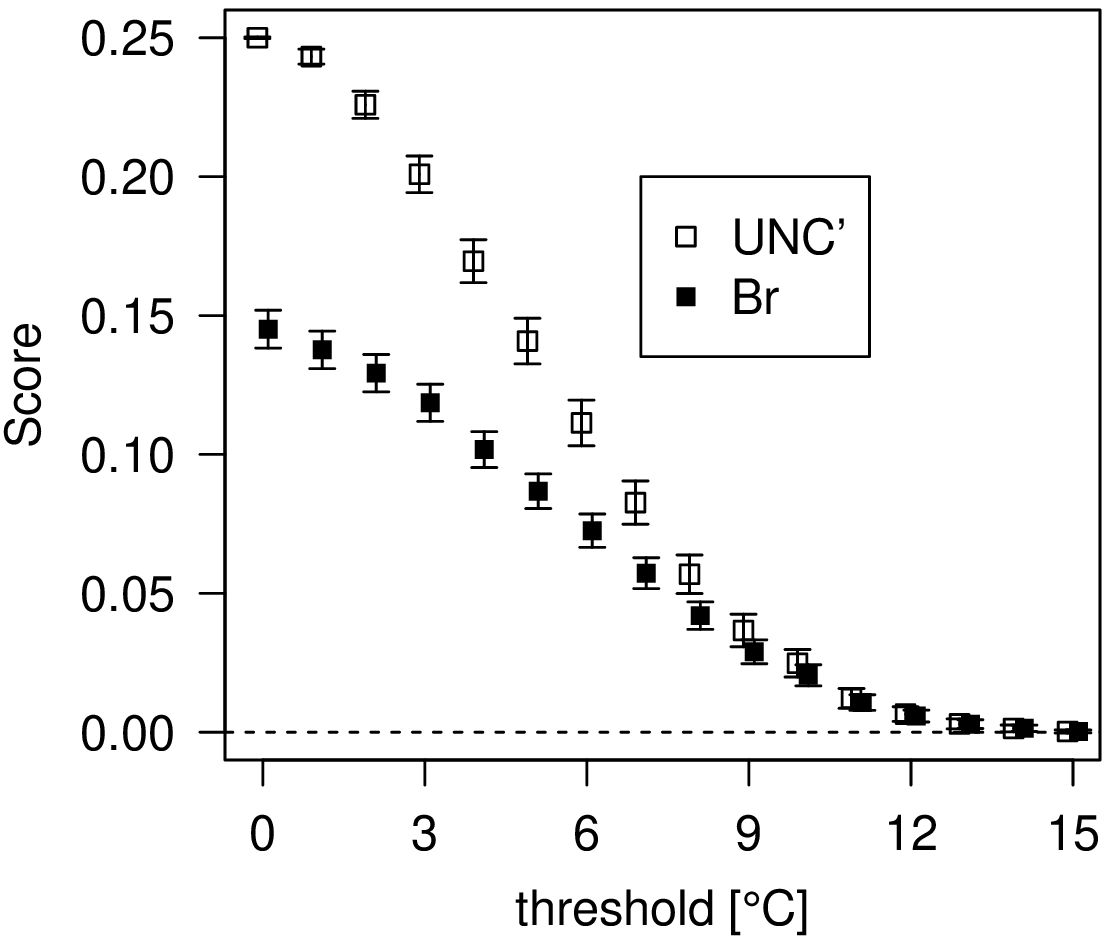}
\caption{Brier Score decomposition of the temperature anomaly exceedance forecasts by an autoregressive model. Upper panel: $\rel'$ and $\res'$ as a function of the threshold which defines the exceedance event, augmented with errorbars of half width two estimated standard deviations. Please note the different y-scales for $\rel'$ and $\res'$. Lower panel: Same as above for $\unc'$ and $\br$. In this plot the y-scale is the same for both quantities.}
\label{fig2}
\end{figure}

In \reffig{fig2}, the bias-corrected components of the autoregressive exceedance forecast and the empirical Brier Score are shown as functions of the threshold.
The error bars of half widths two standard deviations provide an estimate of the sampling variability of the components.
The error bars for the Brier Score were estimated by the standard error of the mean, i.e. $\sqrt{\vv(y-p)^2}/\sqrt{N}$ \citep{bradley2008sampling}.

\section{Discussion}
\label{secDisc}

The assumptions and simplifications that entered the derivation of the variance estimates must be discussed.
The first simplification of the problem was the first order Taylor expansion in \refeq{eqn79}.
Its validity relies on the assumption, that the difference between the observed values of the arguments and their expectation values is small enough that quadratic terms can be ignored.
Furthermore, the covariance matrix that appears in \refeq{eq805} is estimated by the sample covariance.
Both assumptions can lead to errors, especially if $N$, the number of forecasts and verifications is small.
In the light of these criticisms we should expect that more accurate variance estimates than the ones presented here ought to exist.
Nonetheless, \reffig{fig1} suggests that we obtain reasonable variance estimates despite all the simplifying assumptions.
A further simplifying assumption is serial independence: To estimate the covariance matrix by \refeq{eq492}, we made use of the assumption that the pairs of forecast probabilities and event indicators $\{p_n,y_n\}$ are independent for different $n$.
This assumption might not hold in meteorological applications because the probability of rain on day $n+1$, for example, is often similar to the probability of rain on day $n$.
The possible effects serial dependence were not considered at all in this paper.

\refFig{fig1} further shows that the two-standard deviations confidence intervals cover the true value between 91 and 97 out of 100 trials.
For estimators that are Gaussian and unbiased, the expected value would be close to 95.
In the artificial example the biases are about one order of magnitude smaller than the overall variability of the estimators, and the variations of the estimators appear symmetric around their mean and without large deviations.
Unbiasedness and Gaussianity thus seem to be reasonable first approximations to the statistical behavior of the estimators, and the observed coverage frequencies can be taken as further evidence for the quality of the derived variance estimates.
However, coverage should not be overinterpreted as long as the exact statistical properties of the data are not known.

\refTab{tab1} illustrates the decrease of the biases by the estimators derived by \cite{ferro2012unbiased}.
The magnitude of the average difference between the estimator and the true values is substantially lower for the bias-corrected estimators than for the traditional estimators.
At the same time, however, the sample variances of these bias-corrected estimators are slightly larger than the sample variances of the traditional estimators.
This is an example of the bias-variance tradeoff, regularly encountered in statistical estimation problems \citep[e.~g.][]{eldar2008rethinking}.
In fact, \reftab{tab1} shows that the reduction of the bias in the uncertainty, which comes at the cost of an increased variance, leads to a slight increase in the average squared error of this estimator.
That is, even though the bias is reduced, the average squared difference between the estimator and the true value has increased.
For the other two estimators, this is not the case - the increase in variance does not offset the bias-correction.
It should be noted that the increase in sample variance of $\rel'$ is not reflected in the estimated variance which decreases very slightly.
Whether or not any of these effects is systematic cannot be answered at this point and further investigation is required.
The apparent increase in estimator variance is not a generic result, and should certainly not be taken as an argument against the use of bias-corrected estimators.

In \refsec{secExample2} Brier Score decomposition has been applied to autoregressive forecasts of exceedance events of temperature anomalies.
The Brier Score decomposition was applied to 10 years' worth of daily data.
The two-standard-deviation error bars of all estimators are relatively wide, considering that the decomposition is based on more than 3000 data points.
In evaluation studies of weather and climate forecasts, usually much less data is available and the variability of the estimators must be expected to be higher in these cases.
Reliable estimates of the variability of the components of the Brier Score decomposition are required for an honest assessment of the significance of the results.

\section{Summary and conclusions}\label{secSummary}

The components of the Brier Score decomposition can be used to assess the forecast attributes reliability and resolution, as well as the inherent uncertainty of the underlying process.
The decomposition thus provides insight that goes beyond quantifying the performance by calculating the average Brier Score.
Variance estimates were derived for the traditional and bias-corrected estimators of the components of Brier Score decomposition.
The variances were approximated by propagation of uncertainty.
The validity of the variance estimates was illustrated using artificial data, where the true values of the components are known.
An actual meteorological forecast setting illustrated a possible application.
A discussion was provided about the implied assumptions, as well as the consequences of bias-correction.

In the cases considered, the variance estimates provided meaningful approximations as to the statistical variability of the components of Brier Score decomposition.
Confidence intervals exhibited reasonable coverage frequencies, and estimated and empirical variances coincided, despite numerous simplifying assumptions.
Furthermore, it was noted that bias-correction can come at the cost of an increased estimator variance.
An example was shown where the bias-correction was not able to decrease the average squared difference of the estimator from its true value.

Forecasters who want to compare competing probabilistic forecasting schemes based on finite data will certainly find the competing Brier Score components to be different.
Part of this difference is caused by sampling fluctuations.
Using the variance estimates proposed here, the magnitude of these fluctuations can be quantified approximately.
This allows for a more robust comparison in terms of true predictive skill, and helps to avoid overinterpretations of performance differences.

\section*{Acknowledgments}

I thank Colm Mulhern for providing helpful comments on an earlier version of this text.
I am grateful to Jochen Bröcker, Holger Kantz, and Christopher Ferro for fruitful discussions related to the present article, and I acknowledge expert comments from two reviewers which helped greatly to improve the quality of this article.

\begin{appendix}
\section{Appendix: Derivatives}\label{secAppDeriv}

Note that some of the following derivatives can be undefined due to vanishing denominators.
These derivatives must be set to zero.

\subsection{$\rel$}
\begin{align}
\frac{\partial\rel}{\partial A_{\bu d}} & = -\frac{(B_{\bu d}-C_{\bu d})^2}{NA_{\bu d}^2}\\
\frac{\partial\rel}{\partial B_{\bu d}} & = \frac{2(B_{\bu d}-C_{\bu d})}{NA_{\bu d}}\\
\frac{\partial\rel}{\partial C_{\bu d}} & = -\frac{2(B_{\bu d}-C_{\bu d})}{NA_{\bu d}}
\end{align}

\subsection{$\res$}

\begin{align}
\frac{\partial\res}{\partial A_{\bu d}} & = -\frac{1}{N}\left(\frac{B_{\bu d}}{A_{\bu d}}-\frac{Y_\bu}{N}\right)\left(\frac{B_{\bu d}}{A_{\bu d}}+\frac{Y_\bu}{N}\right) \\
\frac{\partial\res}{\partial B_{\bu d}} & = \frac{2}{N}\left(\frac{B_{\bu d}}{A_{\bu d}}-\frac{Y_\bu}{N}\right)\\
\frac{\partial\res}{\partial Y_\bu} & = -\sum_{d\in\mathds{D}_0} \frac{2A_{\bu d}}{N^2}\left(\frac{B_{\bu d}}{A_{\bu d}}-\frac{Y_\bu}{N}\right) \nonumber\\
& = -\frac{2}{N^2}B_{\bu \bu} + \frac{2Y_\bu}{N^3}A_{\bu \bu} = 0\label{eq829}
\end{align}

\subsection{$\unc$}

\begin{equation}
\frac{\partial\unc}{\partial Y_\bu } = \frac{1}{N}-\frac{2Y_\bu }{N^2}
\end{equation}

\subsection{$\rel'$}

\begin{align}
\frac{\partial\rel'}{\partial A_{\bu d}} & = -\frac{1}{NA_{\bu d}^2}\bigg[(B_{\bu d}-C_{\bu d})^2\nonumber\\
&\quad+\frac{B_{\bu d}^2}{A_{\bu d}-1} - \frac{A_{\bu d}B_{\bu d}(A_{\bu d}-B_{\bu d})}{(A_{\bu d}-1)^2}\bigg]\\
\frac{\partial\rel'}{\partial B_{\bu d}} & = \frac{2B_{\bu d}-1}{N(A_{\bu d}-1)} - \frac{2C_{\bu d}}{NA_{\bu d}}\\
\frac{\partial\rel'}{\partial C_{\bu d}} & = -\frac{2(B_{\bu d}-C_{\bu d})}{NA_{\bu d}}
\end{align}

\subsection{$\res'$}

\begin{align}
&\frac{\partial\res'}{\partial A_{\bu d}} = -\frac{1}{N}\left(\frac{B_{\bu d}}{A_{\bu d}}-\frac{Y_\bu }{N}\right)\left(\frac{B_{\bu d}}{A_{\bu d}}+\frac{Y_\bu }{N}\right)\nonumber\\
&\quad + \frac{B_{\bu d}}{NA_{\bu d}^2(A_{\bu d}-1)^2}\left[(A_{\bu d}-B_{\bu d})^2-B_{\bu d}(B_{\bu d}-1)\right]\\
&\frac{\partial\res'}{\partial B_{\bu d}} = \frac{2}{N}\left(\frac{B_{\bu d}}{A_{\bu d}}-\frac{Y_\bu }{N}\right) - \frac{A_{\bu d}-2B_{\bu d}}{NA_{\bu d}(A_{\bu d}-1)}\\
&\frac{\partial\res'}{\partial Y_\bu } = \frac{N-2Y_\bu }{N^2(N-1)}
\end{align}

\subsection{$\unc'$}

\begin{equation}
\frac{\partial\unc'}{\partial Y_\bu } = \frac{N-2Y_\bu }{N(N-1)}
\end{equation}

\section{Appendix: Avoiding inconsistencies due to the bias-correction}\label{secAppAvoid}

Let the variables $S$ and $T$ be defined by $\rel'=\rel-S$ (cf. \refeq{eq317}) and $\unc'=\unc+T$ (cf. \refeq{eq319}).
The bias-correction proposed by \citet{ferro2012unbiased} can be imagined as shifting the 3-vector $\bd = (\rel,\res,\unc)$ to a new point 
\begin{equation}
\bd' = (\rel',\res',\unc') = \bd + \bc,
\end{equation}
where $\bc = (-S, -S+T, T)$, and thus the shift takes place along a plane of constant Brier Score.
Denote by $\mathcal{A} = [0,1]\times [0,1] \times [0,0.25]$ the space of `allowed' Brier Score decompositions.
In order to avoid inconsistencies due to $\bd'\not\in\mathcal{A}$, a possible modification is to use the bias-correction
\begin{equation}
\bd'' = (\rel'',\res'',\unc'') = \bd + \gamma\bc,
\end{equation}
where $\gamma$ is given by
\begin{align}
\gamma = \min\bigg\{ & \frac{\rel}{S},\max\left[\frac{\res}{S-T},\frac{\res-1}{S-T}\right],\nonumber\\
& \frac{1-4\unc}{4T},1\bigg\}.
\end{align}
The parameter $\gamma$ is confined to the unit interval, and ensures that neither $\rel''<0$ nor $\res''<0$ nor $\res''>1$ nor $\unc''>1/4$.
Essentially $\gamma$ ensures that the decomposition $\bd$ is shifted linearly as far as possible to the bias-corrected decomposition $\bd'$, but not too far as to carrying any of the components out of their allowed range.

\end{appendix}

\bibliography{main}

\begin{thebibliography}{13}
\providecommand{\natexlab}[1]{#1}
\providecommand{\url}[1]{\texttt{#1}}
\expandafter\ifx\csname urlstyle\endcsname\relax
  \providecommand{\doi}[1]{doi: #1}\else
  \providecommand{\doi}{doi: \begingroup \urlstyle{rm}\Url}\fi

\bibitem[Bradley et~al.(2008)Bradley, Schwartz, and
  Hashino]{bradley2008sampling}
A.A. Bradley, S.S. Schwartz, and T.~Hashino.
\newblock {Sampling uncertainty and confidence intervals for the Brier score
  and Brier skill score}.
\newblock \emph{Weather and Forecasting}, 23\penalty0 (5):\penalty0 992--1006,
  2008.
\newblock \doi{10.1175/2007WAF2007049.1}.

\bibitem[Brier(1950)]{brier1950verification}
G.W. Brier.
\newblock {Verification of forecasts expressed in terms of probability}.
\newblock \emph{Monthly Weather Review}, 78\penalty0 (1):\penalty0 1--3, 1950.
\newblock \doi{10.1175/1520-0493(1950)078<0001:VOFEIT>2.0.CO;2}.

\bibitem[Bröcker(2008)]{broecker2008remarks}
J.~Bröcker.
\newblock {Some remarks on the reliability of categorical probability
  forecasts}.
\newblock \emph{Monthly Weather Review}, 136\penalty0 (11):\penalty0
  4488--4502, 2008.
\newblock \doi{10.1175/2008MWR2329.1}.

\bibitem[Bröcker(2009)]{broecker2009reliability}
J.~Bröcker.
\newblock {Reliability, sufficiency, and the decomposition of proper scores}.
\newblock \emph{Quarterly Journal of the Royal Meteorological Society},
  135\penalty0 (643):\penalty0 1512--1519, 2009.
\newblock \doi{10.1002/qj.456}.

\bibitem[DeGroot and Fienberg(1983)]{degroot1983comparison}
M.H. DeGroot and S.E. Fienberg.
\newblock {The comparison and evaluation of forecasters}.
\newblock \emph{Journal of the Royal Statistical Society. Series D (The
  Statistician)}, 32\penalty0 (1/2):\penalty0 12--22, 1983.

\bibitem[{Deutscher Wetterdienst}(2012)]{dwd2012}
{Deutscher Wetterdienst}.
\newblock {Climatological Database, online {\tt www.dwd.de} [accessed on 5
  November 2012]}, 2012.

\bibitem[Eldar(2008)]{eldar2008rethinking}
Y.C. Eldar.
\newblock \emph{{Rethinking biased estimation}}, volume~1.
\newblock Now Publishers Inc, 2008.

\bibitem[Ferro and Fricker(2012)]{ferro2012unbiased}
C.A.T. Ferro and T.E. Fricker.
\newblock {A bias-corrected decomposition of the Brier score}.
\newblock \emph{Quarterly Journal of the Royal Meteorological Society},
  138\penalty0 (668):\penalty0 1954--1960, 2012.
\newblock \doi{10.1002/qj.1924}.

\bibitem[Mood et~al.(1974)Mood, Graybill, and Boes]{mood1974introduction}
A.M. Mood, F.A. Graybill, and D.C. Boes.
\newblock \emph{{Introduction to the theory of statistics 3rd ed.}}
\newblock McGraw--Hill, New York, 1974.

\bibitem[Murphy(1973)]{murphy1973new}
A.H. Murphy.
\newblock {A new vector partition of the probability score}.
\newblock \emph{Journal of Applied Meteorology}, 12:\penalty0 595--600, 1973.
\newblock \doi{10.1175/1520-0450(1973)012<0595:ANVPOT>2.0.CO;2}.

\bibitem[{R Core Team}(2012)]{Rmanual2012}
{R Core Team}.
\newblock \emph{{R: A Language and Environment for Statistical Computing}}.
\newblock R Foundation for Statistical Computing, Vienna, Austria, 2012.
\newblock URL \url{http://www.R-project.org/}.
\newblock {ISBN} 3-900051-07-0.

\bibitem[Siegert(2013)]{bride2013}
S.~Siegert.
\newblock {bride: Brier Score decomposition for probabilistic forecasts of
  binary events}.
\newblock R-package (version 1.3), 2013.
\newblock URL \url{http://CRAN.R-project.org/package=bride}.

\bibitem[Stephenson et~al.(2008)Stephenson, Coelho, and
  Jolliffe]{stephenson2008extra}
D.B. Stephenson, C.A.S. Coelho, and Ian~T. Jolliffe.
\newblock {Two extra components in the Brier score decomposition}.
\newblock \emph{Weather and Forecasting}, 23\penalty0 (4):\penalty0 752--757,
  2008.
\newblock \doi{10.1175/2007WAF2006116.1}.

\end{thebibliography}
\end{document}